\documentclass[11pt]{JHEP3} 
\usepackage{epsfig,multicol}
\newcommand{\be}{\begin{equation}}
\newcommand{\ee}{\end{equation}}
\newcommand{\ba}{\begin{eqnarray}}
\newcommand{\ea}{\end{eqnarray}}
\newcommand{\la}{\langle}
\newcommand{\ra}{\rangle}
\newcommand{\di}{d}
\newcommand{\Mn}{M_{\mbox{\tiny N}}}
\newcommand{\M}[1]{M_{\mbox{\tiny ${#1}$}}}
\newcommand{\sigmaPiN}{\sigma_{\mbox{\tiny $\pi$N}}}
\newcommand{\trF}{{\rm tr}_{\mbox{\tiny F}}}
\newcommand{\limNR}{\lim\limits_{
	{\renewcommand{\arraystretch}{0.55}\begin{array}{c}
        \scriptscriptstyle  \rm non \cr
	\scriptscriptstyle  \rm relativistic \end{array}}} \!\!\!\!\!\!  }
\title{The chirally-odd twist-3 distribution \boldmath $e^a(x)$}
\author{A.~V.~Efremov\\
	Joint Institute for Nuclear Research, Dubna, 141980 Russia\\
	E-mail: \email{efremov@thsun1.jinr.ru}}
\author{P.~Schweitzer\\
	Dipartimento di Fisica Nucleare e Teorica, 
  	Universit\`a degli Studi di Pavia, Pavia, Italy\\
	E-mail: \email{peter.schweitzer@pv.infn.it}}

\abstract{
Properties of the nucleon twist-3 distribution function $e^a(x)$ are reviewed.
It is emphasized that the QCD equations of motion imply the existence of a 
$\delta$-function at $x=0$ in $e^a(x)$, which gives rise to the pion-nucleon 
sigma-term. According to the resulting ``practical'' DIS sum rules the first 
and the second moment of $e^a(x)$ vanish, a situation analogue to that of the
pure twist-3 distribution function $\overline{g}_2(x)$.}


\begin{document}

\section{Introduction}

Among the six distribution functions $f_1^a(x)$, $g_1^a(x)$, $h_1^a(x)$ and 
$g_T^a(x)$, $h_L^a(x)$, $e^a(x)$, which describe the structure of the nucleon
in deeply inelastic scattering (DIS) processes up to twist-3, the least known
and studied one is probably  $e^a(x)$ (we use throughout the notation of 
Refs.~\cite{Jaffe:1991kp,Jaffe:1991ra}).
The distribution function $e^a(x)$ is twist-3 and chirally odd. 
Apart from the first moment of $(e^u+e^d)(x)$, which sometimes is related 
to the phenomenologically most interesting pion-nucleon sigma-term, 
it is experimentally unknown.  Only recently it became clear how $e^a(x)$ 
-- in principle -- could be accessed in DIS experiments. 
Most recently the corresponding process has been studied by the HERMES 
and CLAS collaborations. In particular, the CLAS data possibly provide 
the first experimental indications for $e^a(x)$.

This note has 
partly the character of a brief review, 
however, also some new results are reported.
In Sec.~\ref{Sec:e-in-theory} the definition of $e^a(x)$ is given,
and its theoretical properties are discussed. 
The known but only casually mentioned fact is emphasized, 
that $e^a(x)$ contains a $\delta(x)$-contribution.
Different statements in literature on the small-$x$ behaviour of $e^a(x)$, 
which at first glance seem to be contradictory, are shown to be consistent.
Sum rules for $e^a(x)$ are discussed. 
It is argued that there is no twist-3 inequality constraining
$e^a(x)$ in terms of other twist-3 distribution functions.
In Sec.~\ref{Sec:e-in-models} a brief overview is given about model 
calculations of $e^a(x)$. In particular results from the non-relativistic 
model, bag model, spectator model and chiral quark-soliton model are discussed.
Sec.~\ref{Sec:e-in-experiments} briefly reports the recent progress on 
understanding time-odd phenomena -- in particular in the fragmentation 
processes -- which give rise to single spin asymmetries.
Such asymmetries have recently been studied by the HERMES and CLAS 
collaborations. 
Sec.~\ref{Sec:conlusions} contains the summary and conclusions.
Some technical details concerning the gauge invariant decomposition 
of $e^a(x)$ can be found in App.~\ref{App:using-eq-of-motion}.

\section{\boldmath $e^q(x)$ in theory}
\label{Sec:e-in-theory}

\paragraph{Definition.}
The chirally odd twist-3 distribution functions $e^q(x)$ and $e^{\bar q}(x)$
for quarks of flavour $q$ and antiquarks of flavour $\bar q$ 
are defined as \cite{Jaffe:1991kp,Jaffe:1991ra}
\be\label{definition}
	e^q(x) = \frac{1}{2\Mn} \int\!\frac{\di\lambda}{2\pi}\,e^{i\lambda x}\,
	\la N|\,\bar{\psi}_q(0)\,[0,\lambda n]\,\psi_q(\lambda n)\,|N\ra 
	\;,\;\;\; e^{\bar q}(x)=e^q(-x) \;,\ee
where $[0,\lambda n]$ denotes the gauge-link. 
The scale dependence is not indicated for brevity. 
The light-like vectors $n^\mu$ in Eq.~(\ref{definition}) and $p^\mu$ are 
defined  such that $n^\mu p_\mu=1$ and the nucleon momentum is given by  
$P^\mu_{\!\!\mbox{\tiny N}}=n^\mu+ \frac12 \Mn^2 p^\mu$.
The matrix element in Eq.~(\ref{definition}) is averaged over nucleon spin,
i.e. $\la N|\dots|N\ra \equiv \frac12\sum_{S_3}\la N,S_3|\dots|N,S_3\ra$.

\paragraph{Evolution.}
The renormalization scale dependence of $e^a(x)$ has been studied 
in Refs.~\cite{Balitsky:1996uh,Belitsky:1997zw,Koike:1997bs},
see also Refs.~\cite{Belitsky:1997ay,Kodaira:1998jn} for reviews.
The evolution of $e^a(x)$ is characterized by a complicated 
operator mixing pattern typical for twist-3 quantities.
In the multi-colour limit the evolution of $e^a(x)$ simplifies to a
DGLAP-type evolution -- as it does for the other two nucleon twist-3 
distribution functions $h_L^a(x)$ and (the flavour non-singlet) $g_T^a(x)$.

\paragraph{\boldmath Sum rules for the $1^{\rm st}$ and $2^{\rm nd}$ moment.}
The first moment of $(e^u+e^d)(x)$ is related to the pion-nucleon sigma-term 
\cite{Jaffe:1991ra}
\be\label{1moment}
	\int_{-1}^1\di x\;(e^u+e^d)(x)  =  \frac{1}{2\Mn}\la N|\,
	\left(\bar{\psi}_u\psi_u+\bar{\psi}_d\psi_d\right)\,|N\ra
	\equiv \frac{\sigmaPiN}{m} \;\;, \ee
$m \equiv\frac12\,(m_u+m_d)$ is the average mass of the light quarks.
The pion-nucleon sigma-term $\sigmaPiN$ \cite{sigma-review} is defined 
as the value of the nucleon scalar isoscalar form factor $\sigma(t)$,
\be\label{sigma(t)}
	\sigma(t) = \frac{m}{2\Mn\;} \la N(P')|\,
	\left(\bar{\psi}_u\psi_u+\bar{\psi}_d\psi_d\right)\,|N(P)\ra 
	\;\;,\;\; t = (P-P')^2 \;\;,\ee
at $t=0$, i.e. $\sigmaPiN\equiv\sigma(0)$.\footnote{
	More precisely $\sigma(t)$ is defined as the nucleon form factor
	of the double commutator of the strong interaction Hamiltonian with 
	two axial isovector charges \cite{sigma-review}. In the definition 
	(\ref{sigma(t)}) a ``double isospin violating term'' proportional to
	$(m_u-m_d)(\bar{\psi}_u\psi_u-\bar{\psi}_d\psi_d)$ is neglected.}
The relation (\ref{1moment}) of $e^q(x)$ to $\sigmaPiN$ is correct in a formal 
mathematical sense. The sum rule (\ref{1moment}), however, unfortunately 
is of {\sl no practical use} (not even in principle) to gain information 
on $\sigmaPiN$ from DIS experiments -- as we shall see below.

The pion-nucleon sigma-term $\sigmaPiN$
gives the amount by which the nucleon mass changes, 
when the $u$- and $d$-quarks are given a small mass $m$ \cite{sigma-review}.
The form factor $\sigma(t)$ describes the elastic scattering off the nucleon 
due to the exchange of an isoscalar spin-zero particle, and is not known 
experimentally except for its value at the Cheng-Dashen point $t=2m_\pi^2$.
Low energy theorems \cite{low-energy-theorem} allow to relate 
$\sigma(2m_\pi^2)$ to pion nucleon scattering amplitudes and one finds
\be\label{sigma(2m_pi^2)}
	\sigma(2m_\pi^2) = \cases{
	(64\pm 8)\,{\rm MeV} & Ref.~\cite{Koch:pu}\cr
	(79\pm 7)\,{\rm MeV} & Ref.~\cite{Pavan:2001wz}.}\ee
The difference $\sigma(2m_\pi^2)-\sigma(0)$ has been calculated from 
a dispersion relation analysis \cite{Gasser:1990ce} and in chiral 
perturbation theory \cite{Becher:1999he} with the consistent result
\be\label{sigma-diff}
	\sigma(2m_\pi^2)-\sigma(0) \simeq 14\;{\rm MeV} \;.
\ee
This means that 
\be\label{sigmaPiN}
	\sigmaPiN = \sigma(0) \simeq (50-65)\,{\rm MeV} \;. \ee
With $m \simeq 7 \,{\rm MeV}$ (at a scale of say $1\,{\rm GeV}^2$) 
one concludes a large number for the first moment of $(e^u+e^d)(x)$
\be\label{1moment-number}
	\int_{-1}^1\di x\;(e^u+e^d)(x) \simeq (6-10) \;. \ee
Since $\sigmaPiN$ is normalization scale invariant while the running 
current quark masses decrease with increasing scale, the number in 
Eq.~(\ref{1moment-number}) becomes even larger at higher scales.

The second moment of $e^q(x)$ is proportional to the number of the 
respective valence quarks $N_q$ (for proton $N_u=2$ and $N_d=1$) 
and vanishes in the chiral limit \cite{Jaffe:1991ra}
\be\label{2moment}
	\int_{-1}^1\di x\;x\,e^q(x) = \frac{m_q}{\Mn}\; N_q \;. \ee

\paragraph{Twist-3 inequality.}
In Ref.~\cite{Soffer:1995ww} the ``Soffer-inequality'' for twist-2
nucleon distribution functions,
\be\label{Sof-twist-2}
	f_1^a(x) + g_1^a(x) \ge 2 |h_1^a(x)| \;\;, \ee
was obtained making use of the positivity of the scattering density matrix. 
(An alternative derivation was given shortly after in \cite{Goldstein:1995ek}.)
Assuming that this argument can be generalized to higher twists,
similar inequalities were obtained for twist-3 and twist-4 distribution 
functions in \cite{Soffer:1995ww}. In particular, the ``twist-3 Soffer 
inequality'' reads $e^a(x) + h_L^a(x) \ge 2 |g_T^a(x)|$. Unfortunately, 
in general the positivity argument is not valid for higher twists.

One way to understand this is to recall that the twist-2 inequalities are 
derived by relating the imaginary part of the elastic forward quark-nucleon
scattering amplitude by means of the optical theorem to the total cross
section (which is positive). In the Bjorken-limit the imaginary part of the
amplitude can be expressed in terms of twist-2 parton distribution functions.
Twist-3 effects, of course, can be taken into account. They appear as
corrections of the order $\Mn/Q$ to the imaginary part of the amplitude.
There is, however, in general no reason for such corrections to be positive. 
In other words, it is not possible to impose positivity at the level of each 
twist separately, fore the positivity of the scattering density matrix is 
already guaranteed by the twist-2 distribution functions (in the limit of 
large $Q\gg \Mn$).
Interestingly, if for some reason the twist-2 inequality (\ref{Sof-twist-2}) 
is saturated, i.e.\ if $f_1^a(x)+g_1^a(x)=2|h_1^a(x)|$, then the positivity 
of the cross section requires certain twist-3 inequalities to hold, and in
one case $e^a(x)+h_L^a(x)\ge 2|g_T^a(x)|$ holds.  

A different argument why twist-3 inequalities generally fail was given in
Ref.~\cite{Goldstein:1995ek}. In the framework of light-cone formalism it
was demonstrated that possible positivity constraints on the twist-3 
distributions $e^a(x)$, $h_L^a(x)$ and $g_T^a(x)$ inevitably involve
twist-2 {\sl and} twist-4 distributions -- which makes such constraints
practically useless \cite{Goldstein:1995ek}.  (Still, in some cases 
useful constraints -- though involving different twists -- can be
obtained, see \cite{Soffer:2000zd}.)

\paragraph{\boldmath The large $N_c$ limit.}
In the limit of a large number of colours $N_c$ one observes the following 
behaviour of the singlet and non-singlet flavour combinations 
\cite{Schweitzer:2003uy}
\ba\label{large-Nc-behaviour-in-general}
&&	(e^u+e^d)(x) = N_c^2 d(N_cx) \nonumber\\
&&	(e^u-e^d)(x) = N_c\, d(N_cx) \;, \ea
where the functions $d(y)$ are stable in the large-$N_c$ limit for a fixed 
argument $y=N_cx$, and of course different for the different flavour 
combinations. From (\ref{large-Nc-behaviour-in-general}) we conclude that 
\be\label{large-Nc-2}
	|(e^u+e^d)(x)| \gg |(e^u-e^d)(x)| \;\;\;\mbox{for $N_c\to\infty$,}\ee
or in other words $e^u(x) \approx e^d(x)$ modulo $1/N_c$-corrections.
Such large-$N_c$ relations hold well in nature, 
see e.g.\ Ref.~\cite{Efremov:2000ar}, and can serve as a useful guideline.

\paragraph{\boldmath 
Decomposing $e^a(x)$ by means of the QCD equations of motion.}
The following operator identity follows from the QCD-equations of motion 
(flavour indices on the quark fields are omitted for simplicity)
\ba\label{identity}
&&\!\!\!\!\!\!\!\!\!\!\!\!\!\!\!\!
	\bar\psi(0)\,[0,z]\,\psi(z) 
	= \bar\psi(0)\,\psi(0) \nonumber\\
&&\;\;	+\; \frac12\,\int\limits_0^1\di u\int\limits_0^u\di v\;
	    \bar\psi(0)\,\sigma^{\alpha\beta}z_\beta\,[0,vz]\,
	    gG_{\alpha\nu}(vz)\,z^\nu\,[vz,uz]\,\psi(uz) \nonumber\\
&&\;\;	-\; i m_q \int\limits_0^1\di u\; 
 	    \bar\psi(0)\not{\!z}\,[0,uz]\,\psi(uz) \nonumber\\
&&\;\;	-\; \frac{i}{2} \int\limits_0^1\di u \biggl(
	    \bar\psi(0)\,(i\!\not{\!\!D}-m_q)\not{\!z}\,[0,uz]\,\psi(uz)
	    +\bar\psi(0)\not{\!z}\,[0,uz]\,(i\!\not{\!\!D}-m_q)\,\psi(uz)
	    \biggr) \;.\nonumber\\ \;\;\;\;\;\;\ea
The identity (\ref{identity}) is exact up to total derivatives which are 
irrelevant for the parton distribution functions. The formalism to derive 
such identities has been introduced in Ref.~\cite{Braun}. 
The identity (\ref{identity}) can be found, e.g., in 
Refs.~\cite{Belitsky:1997zw,Kodaira:1998jn}.
The ``equations of motion'' operator in the last line of Eq.~(\ref{identity})
vanishes in physical matrix elements, however, its mixing under 
renormalization with the other operators in Eq.~(\ref{identity}) has to 
be considered in the study of the evolution properties of $e^a(x)$, see 
\cite{Balitsky:1996uh,Belitsky:1997zw,Koike:1997bs} and 
\cite{Belitsky:1997ay,Kodaira:1998jn}.
The other operators in the identity (\ref{identity}),
after inserted into the definition (\ref{definition}),
yield the following decomposition of $e^q(x)$
\be\label{e-decompostion-1}
	e^q(x) =  e^q_{\rm sing}(x)
	       	+ e^q_{\rm tw3}(x)
		+ e^q_{\rm mass}(x) 	\;\;.\ee
The contributions $e^q_{\rm sing}(x)$, $e^q_{\rm tw3}(x)$ and 
$e^q_{\rm mass}(x)$ are ``physically real'', in the sense that each term 
in the operator decomposition in Eq.~(\ref{identity}) is gauge-invariant. 
They are defined as follows (see also App.~\ref{App:using-eq-of-motion}).

The contribution $e^q_{\rm sing}(x)$ arises from the local scalar operator 
$\bar\psi_q(0)\psi_q(0)$ on the right-hand-side (RHS) of the identity 
(\ref{identity}). It is proportional to a $\delta$-function at $x=0$
\be\label{e-decompostion-sing}
	e^q_{\rm sing}(x) = \frac{\delta(x)}{2\Mn}\,
	\la N|\bar\psi_q(0)\psi_q(0)|N\ra \;. \ee
The presence of this singular term is well known but only rarely mentioned
(see, e.g., the footnote on p.~233 of Ref.~\cite{Kodaira:1998jn}).
It is customary to cancel out this contribution by multiplying $e^q(x)$ by 
$x$, since it has no partonic interpretation and is not relevant for the
discussion of the evolution properties of $e^q(x)$. However, this contribution
gives rise to the pion-nucleon sigma-term \cite{Balitsky:1996uh}.
The possible existence of a $\delta(x)$ singularity in $e^q(x)$ 
and structure functions associated with it has a long history.
We shall come to this point later on.

The contribution $e^q_{\rm tw3}(x)$ is a quark-antiquark-gluon correlation 
function, i.e. the actual ``pure'' twist-3, ``interaction dependent'' 
contribution to $e^q(x)$. It has a ``partonic interpretation'' as an 
interference between scattering from a coherent quark-gluon pair and from 
a single quark, see \cite{Jaffe:1991kp,Jaffe:1991ra} and references therein. 
It is due to the second operator on the RHS of the identity (\ref{identity}).
The explicit expression of this term can be found in 
\cite{Jaffe:1991ra,Balitsky:1996uh,Belitsky:1997zw,Koike:1997bs} 
and \cite{Belitsky:1997ay,Kodaira:1998jn} 
(see also App.~\ref{App:using-eq-of-motion}).
Here we only mention that the first two moments of 
$e^q_{\rm tw3}(x)$ vanish, i.e.
\be\label{e-decompostion-twist3-mom12}
	{\cal M}_n[e^q_{\rm tw3}] 
	= \int\limits_{-1}^1\!\di x\;x^{n-1}\,e^q_{\rm tw3}(x)
	= 0 \;\;\; \mbox{for $n=1,\,2$.} \ee

The contribution $e^q_{\rm mass}(x)$ is proportional to the current 
quark mass, and is conveniently defined in terms of its Mellin moments
\be\label{e-decompostion-mass-mom}
	{\cal M}_n[e^q_{\rm mass}] = \frac{m_q}{\Mn}\times
	\cases{0                    & for $n=1$,\cr
	      {\cal M}_{n-1}[f_1^q] & for $n>1$,}\ee
where ${\cal M}_n[f_1^q]$ are the moments of the twist-2 unpolarized
distribution $f_1^q(x)$. The relation (\ref{e-decompostion-mass-mom})
for moments $n\ge 2$ can be inverted as 
\be\label{e-decompostion-mass}
	x\,e^q_{\rm mass}(x) 
	= \frac{m_q}{\Mn}\; f_1^q(x) \;.\ee
Of course, for $x\neq 0$ one can divide Eq.~(\ref{e-decompostion-mass})
by $x$  --  this allows to draw conclusions on the behaviour of $e^q(x)$ at 
small $x>0$ (see below). However, one has to keep in mind that the correct 
definition of $e^q_{\rm mass}(x)$ is Eq.~(\ref{e-decompostion-mass-mom}).

\paragraph{\boldmath Small-$x$ behaviour of $e^q(x)$.}
The small- and large-$x$ behaviour of pure twist-3 distributions has been 
studied, see e.g.\ Eqs.~(6.38, 6.39) in Ref.~\cite{Kodaira:1998jn}.
For the following discussion we note the result
\be\label{e-twist3-small-x}
	e^q_{\rm tw3}(x) \to {\rm const}\;x^0 \;\;\;
	\mbox{for $x\to0$.} \ee
Let us consider small $x\neq 0$ and the chiral limit ($m_q\to0$), 
where the mass-term in Eq.~(\ref{e-decompostion-mass}) drops out.
Then Eq.~(\ref{e-twist3-small-x}) dictates the small-$x$ behaviour of 
$e^q(x)$, and we have $e^q(x)\to {\rm const}$ for small but non-zero $x$.
This result is consistent with the conclusion 
\be\label{small-x-Efremov}
	e^q(x)\to {\rm const}\;x^{-0.04}\;\;\;\mbox{for $x\to0$,} \ee
drawn in Ref.~\cite{Efremov:2002ut} from Regge phenomenology assuming a
linear trajectory with standard slope (because the Regge trajectory 
could be slightly non-linear or its slope could be slightly different 
from $1\,{\rm GeV}^{-2}$ and have an intercept $\alpha(0)=-1$). 
In particular, in \cite{Efremov:2002ut} it was concluded that the 
Pomeron decouples, because the Pomeron residue is spin-non-flip.

Considering finite quark mass effects, however, we see that the behaviour 
of $e^q(x)$ for small but non-zero $x$ is dominated by the mass-term 
$e^q_{\rm mass}(x) = m_q f_1^q(x)/(\Mn x)$ in 
Eq.~(\ref{e-decompostion-mass}), and that the Pomeron contributes to $e^q(x)$ 
(since it contributes to $f_1^q(x)$). This agrees with the conclusion
\be\label{small-x-Jaffe+Ji}
	e^q(x) \to {\rm const}\;\frac{1}{x^2}\;\;\;\mbox{for $x\to0$,} \ee
drawn in Ref.~\cite{Jaffe:1991ra} from Regge phenomenology, where it was 
argued that the ``Pomeron couples'', even though suppressed by the factor 
$m_q/\Mn$.

The small $x$ behaviour of $e^q(x)$ in (\ref{small-x-Jaffe+Ji}) makes -- at 
first glance -- questionable the convergence of the sum rule (\ref{1moment})
\cite{Jaffe:1991ra}. However, from Eq.~(\ref{e-decompostion-mass-mom}) 
we see that $e^q_{\rm mass}(x)$ does not contribute to (\ref{1moment}). 
Thus it is not the small $x$ behaviour in (\ref{small-x-Jaffe+Ji})
which makes the sum rule (\ref{1moment}) practically useless for the
purpose of relating $e^q(x)$, as it could be measured in DIS, to the pion 
nucleon sigma term. Rather it is, of course, the $\delta(x)$ contribution.

\paragraph{Conclusions from the use of equations of motion.}
The first conclusion is that the pion nucleon sigma term originates from the 
singular $\delta(x)$-contribution $e^q_{\rm sing}(x)$ only. This 
follows from comparing Eqs.~(\ref{1moment}) and (\ref{e-decompostion-sing})
and considering 
Eqs.~(\ref{e-decompostion-twist3-mom12},\ref{e-decompostion-mass-mom}), i.e.
\be\label{conclusion-eom-1}
	\int\limits_{-1}^1\!\di x\;e^q(x) = 
	\int\limits_{-1}^1\!\di x\;e^q_{\rm sing}(x) = 
	\frac{1}{2\Mn}\,\la N|\bar\psi_q(0)\psi_q(0)|N\ra \;.\ee
If one neglects current quark mass effects\footnote{
	Finite quark mass effects are considered in the next paragraph.}
Eq.~(\ref{conclusion-eom-1}) has the following consequence.
Recalling that $e^{\bar q}(x)=e^q(-x)$ and integrating over $x$ in the 
interval $[0_+,1] \equiv [\epsilon,1]$ with a positive $\epsilon$ 
arbitrarily close (but not equal) to zero, one obtains 
\be\label{conclusion-eom-2}
	\int\limits_{0+}^1\!\di x\;(e^q+e^{\bar q})(x) = 0 \;\; . \ee
The existence of the $\delta(x)$, of course, cannot be confirmed in the
experiment. Eq.~(\ref{conclusion-eom-2}), however, corresponds to the
experimental situation and could in principle be tested in the experiment.

\paragraph{Finite current quark mass effects.}
Neither the pure twist-3 contribution $e^q_{\rm tw3}(x)$ 
(due to Eq.~(\ref{e-decompostion-twist3-mom12})) nor the singular term 
$e^q_{\rm sing}(x)$ (due to $\int_{-1}^1\di x\;x\,\delta(x)=0$)
contribute to the second moment of $e^q(x)$. Thus the sum rule in 
Eq.~(\ref{2moment}) is saturated by the mass term, i.e.
\be
	\int\limits_0^1\!\di x\;x\,(e^q-e^{\bar q})(x) = \frac{m_q}{\Mn} 
	\int\limits_0^1\!\di x\;(f_1^q-f_1^{\bar q})(x) = 
	\frac{m_q}{\Mn}\,N_q \;. \ee

Let us investigate in detail the effect of finite quark mass in 
Eq.~(\ref{conclusion-eom-2}). The integral of $e^a_{\rm mass}(x)$ over $x$ 
in $[-1,1]$ yields zero according to Eq.~(\ref{e-decompostion-mass-mom}).
So for any $0<x_{\rm min}\le 1$ the following equation is formally true
\be\label{App2-1}
	\int\limits_0^{x_{\rm min}}\!\!\di x\; (e^q_{\rm mass}
	+e^{\bar q}_{\rm mass} )(x) 
	= - \int\limits_{x_{\rm min}}^1\!\!\di x\; (e^q_{\rm mass}
	  + e^{\bar q}_{\rm mass} )(x) 
 	= - \frac{m_q}{\Mn} \int\limits_{x_{\rm min}}^1\!\di x\;
	\frac{(f_1^q+f_1^{\bar q})(x)}{x} \le  0 \;.\ee
In the third integral in (\ref{App2-1}) we made use of 
Eq.~(\ref{e-decompostion-mass}) divided by $x$ (which is allowed since the 
point $x=0$ is not included in that integral). The final step in (\ref{App2-1})
follows from the positivity of $f_1^a(x)$. All integrals in (\ref{App2-1}) 
are (formally) well defined for any $x_{\rm min}>0$. 

Thus, if one does not neglect $m_q/\Mn$,  the DIS sum rule 
(\ref{conclusion-eom-2}) for the first moment of $e^a(x)$ becomes\footnote{
	Due to (\ref{e-decompostion-twist3-mom12}) we have
	$\int_{x_{\rm min}}^1\di x\,(e^q_{\rm tw3}+e^{\bar q}_{\rm tw3})(x)=$
	$-\int_0^{x_{\rm min}}\di x\,(e^q_{\rm tw3}+e^{\bar q}_{\rm tw3})(x)$
	$\approx 0$ for small $x_{\rm min}$ because of the smooth behaviour 
	of $e^a_{\rm tw3}(x)$ at small $x$ in (\ref{e-twist3-small-x}).
	This allows to safely neglect the contribution from $e^q_{\rm tw3}(x)$
	to (\ref{App2-2}).}
\be\label{App2-2}
	\int\limits_{x_{\rm min}}^1\!\!\!\di x\;(e^q+e^{\bar q})(x)
	= \frac{m_q}{\Mn}\int\limits_{x_{\rm min}}^1\!\!\!\di x\;
	\frac{(f_1^q+f_1^{\bar q})(x)}{x} 
	\;\;\;\mbox{for very small $x_{\rm min}>0$.} \ee
It is clear that also $e^a_{\rm mass}(x)$ contains a singularity at 
$x=0$. Formally this singularity can be written as a generalized distribution
\be\label{App2-3}
	e^q_{\rm mass}(x) 
	= \frac{m_q}{\Mn}\biggl[\,{\cal P}\,\frac{f_1^q(x)}{x} -\delta(x)\;
	{\cal P}\int\limits_{-1}^1\!\!\di x'\;\frac{f_1^q(x')}{x'}\,\biggr]
	\stackrel{\rm or}{=}
	\lim\limits_{\epsilon\to 0} \frac{m_q}{\Mn}
	\biggl[\frac{x f_1^q(x)}{x^2+\epsilon^2} - \delta(x)\int\limits_{-1}^1
	\!\!\di x'\;\frac{x'f_1^q(x')}{{x'}^2+\epsilon^2}\, \biggr] \;, \ee
where it is understood that the principal value prescription (or the limit 
$\epsilon\to 0$) has to be taken only after $e^q_{\rm mass}(x)$ has been 
inserted in an integral and integrated over. Thus -- for finite $m_q$ -- 
there is formally yet another $\delta$-function at $x=0$ in $e^a(x)$. 
This $\delta$-function ensures the formal ``convergence'' 
(when the point $x=0$ is included) of the sum rule 
(\ref{1moment}) by cancelling the contribution from the mass term 
which strongly rises with decreasing $x$, cf.\ Eq.~(\ref{small-x-Jaffe+Ji}).
Thus the paradoxical situation emerges that the sum rule (\ref{1moment})
``practically'' (since $x=0$ cannot be reached experimentally) diverges,
as noticed in \cite{Jaffe:1991ra}.  But ``theoretically'' (when the 
point $x=0$ is included) the sum rule (\ref{1moment}) exists  
(and is then saturated by the $\delta(x)$ 
contribution in Eq.~(\ref{e-decompostion-sing})).

In principle, the small factor $m_q/\Mn$ in Eq.~(\ref{App2-2}) could be 
compensated by the factor 
$\int_{x_{\rm min}}^1\di x\,(f_1^q+f_1^{\bar q})(x)/x$
which rapidly grows with decreasing $x_{\rm min}$.  Does this mean that the 
relation (\ref{App2-2}) could in principle be used to measure current quark 
masses in DIS? 
At leading twist, current quark mass effects are not observable in DIS 
because they are suppressed by a hard power $m_q/Q$ and cannot be 
distinguished from other (possibly non-factorizing) power suppressed 
contributions which are generically ${\cal O}(\Lambda_{\rm QCD}/Q)$. 
The attempt to ``measure'' $m_q$ by means of Eq.~(\ref{App2-2}) is also
of such kind: Presuming factorization the physical contribution to an 
observable of the twist-3 $e^a(x)$ is accompanied by the factor $\Mn/Q$,
i.e.\ the effect of $m_q$ is effectively $(\Mn/Q)\times(m_q/\Mn) = m_q/Q$.
The (purely academic) question, whether $m_q$ could be measured in this way
in DIS, would be answered by thorough proofs of factorization for processes 
involving $e^a(x)$. Such proofs would clarify whether $m_q$-contributions 
factorize from infrared singularities (in a process-independent way). 
At present, no such proof exists. 

It is interesting to remark that -- wherever it was {\sl assumed}
that factorization holds \cite{muldt,Levelt:1994np,Mulders:1996dh} 
-- it was always $x\,e^a(x)$ (and not $e^a(x)$) which contributed 
to the cross-section. From a practical point of view, one can thus 
redefine $e^a(x)\to e^a_{\rm red}(x)\equiv x\,e^a(x)$ 
(as indeed some authors \cite{Belitsky:1997zw,Belitsky:1997ay} do),
and the discussions about $\delta(x)$-functions become superfluous. 
However, then one faces the interesting phenomenon, that the 
pion-nucleon sigma-term originates from a non-physical -- namely 
the ``minus first'' -- moment of the redefined $e^a_{\rm red}(x)$.
The continuation to non-physical negative moments has been 
discussed in Ref.~\cite{Bukhvostov:rn}.

\paragraph{\boldmath The $\delta(x)$ singularity in $e^q(x)$.}
We have seen that QCD equations of motion allow to decompose $e^q(x)$ in a 
gauge invariant way into three contributions, one of them being proportional 
to a $\delta$-function type singularity at $x=0$. As the limiting point $x=0$
can neither be attributed to quark nor to antiquark distribution functions 
this finding is more clearly expressed as
\be\label{Eq:new}
	(e^q+e^{\bar q})(x) = 
	\frac{\delta(x)}{2\Mn}\,\la N|\bar\psi_q(0)\psi_q(0)|N\ra
	+ \mbox{regular pure twist-3} 
	+ {\cal O}\left(\frac{m_q}{\Mn}\right) \;. \ee
This means that the connection of $e^q(x)$ and the pion nucleon sigma term 
is of purely formal character; there is no experimental relation between
$\sigmaPiN$ and $e^q(x)$ or a structure function related to it.
Interestingly, this possibility was considered already in the early
1970s before the advent of QCD \cite{Jaffe:1972yv}.

In order to carefully derive DIS sum rules -- such as (\ref{1moment}) -- one 
uses dispersion relations to relate the (at least in principle) measurable 
structure functions to the imaginary part of the respective forward scattering
amplitude. The latter can then further be investigated by means of the 
operator product expansion, which in the Bjorken limit allows to connect 
moments of the structure functions to matrix elements of local operators.
A sum rule formally derived using the operator product expansion is valid 
also for the experimentally measurable structure function if the forward 
scattering amplitude satisfies {\sl unsubtracted} dispersion relations. 
If subtraction terms -- in the context of Regge phenomenology referred to
as ``fixed poles'' \cite{Brodsky:1971zh} -- have to be included to make 
the dispersion integral finite, then the sum rule can be spoiled.
A subtraction term in the dispersion integral manifests itself as a 
$\delta(x)$-contribution in the structure function \cite{Broadhurst:1973fr}.
(Cf.\  \cite{Burkardt:1995ts} for a nice pedagogical exposition.)
On the basis of such dispersion relation and Regge arguments it was 
observed in Ref.~\cite{Jaffe:1972yv} that the sigma term sum rule\footnote{
	More precisely, it is the sum rule involving the structure 
	functions $F_4(x)$ and $F_5(x)$ which are related to $e^q(x)$ 
	and -- in principle -- could be measured in 
	(anti)neutrino-nucleon DIS.} 
(\ref{1moment}) could be spoiled.
More prominent examples of sum rules which could possibly be spoiled in 
this way are the Burkhardt-Cottingham sum rule \cite{Burkhardt:ti} and 
the Gerasimov-Drell-Hearn sum rule \cite{Drell:1966jv}.

Further indications towards the presence of a $\delta(x)$-contribution in 
$e^q(x)$ were presented in Ref.~\cite{Burkardt:2001iy}, where $e^q(x)$ was 
constructed explicitly for a one-loop dressed massive quark. Of course, the 
perturbative calculation of Ref.~\cite{Burkardt:2001iy} does not {\sl prove} 
the existence of a $\delta(x)$-contribution in $e^q(x)$. 
But it {\sl strongly suggests} it since one hardly can imagine a mechanism 
to cancel a $\delta(x)$-contribution, which appears in the leading 
order of some small coupling expansion, by higher order contributions.
In the next section we shall see that a $\delta(x)$-contribution has 
recently been observed also in {\sl non-perturbative} calculations 
in the framework of a realistic model of the nucleon
(chiral quark-soliton model).

\section{\boldmath $e^q(x)$ in models}
\label{Sec:e-in-models}

In this section we review results obtained in the non-relativistic quark 
model, bag model, spectator model and chiral quark-soliton model. 
We also will mention calculations in some toy models.

A subtle question is 
whether twist-3 distribution function can be described in models with no gluon
degrees of freedom. However, among the most general twist-3 structures in the 
nucleon $e^a(x)$, $g_T^a(x)$ and $h_L^a(x)$ are distinguished inasmuch they 
can be expressed in terms of quark fields only, i.e.\ with no {\sl explicit} 
gluon fields. This allows to describe these distribution functions in models
with no gluon degrees of freedom, as argued in 
\cite{Jaffe:1991kp,Jaffe:1991ra}. 
However, the results of such model calculations have to be interpreted
with care.

\paragraph{Non-relativistic quark model.}
The non-relativistic limit is an intuitive and, in some cases, useful 
guideline.  We recall the popular relation $h_1^q(x)=g_1^q(x)$, which is 
often used to estimate effects of the transversity distribution function. 
(Irrespective the fact that, taken literally, the non-relativistic model yields
$h_1^q(x)=g_1^q(x)=P_q\delta(x-\frac13)$ with $P_u=4/3$ and $P_d=-1/3$.) 
In this paragraph $q=u,\,d$ since there are no antiquarks in the 
non-relativistic limit, and $m_q=m_u=m_d$ is to be understood as the 
constituent mass of the light quarks, which is one third of the nucleon mass, 
i.e.\ $m_q=\Mn/3$.

In the non-relativistic limit the twist-3 quark distribution $e^q(x)$ and 
the unpolarized twist-2 quark distribution $f_1^q(x)$ coincide 
\cite{Schweitzer:2003uy}
\be\label{non-rel-limit}
	\limNR e^q(x) = \limNR f_1^q(x) = N_q\;\,\delta\!\left(x-\frac13\right)
	\;. \ee
For the first moment the result in (\ref{non-rel-limit}) yields
$\int\di x\,(e^u+e^d)(x)=3$. This is the correct non-relativistic result 
for the sum rule (\ref{1moment}), since in this limit $\sigmaPiN=3m_q=\Mn$.
The latter can be verified by taking the non-relativistic limit in the 
expression (\ref{sigma(t)}) for $\sigmaPiN$, or alternatively 
by means of the Feynman-Hellmann theorem 
\be\label{sigma-Mn-relation}
        \sigmaPiN = m \;\frac{\partial\Mn(m)}{\partial m}\;,\ee
where $m=\frac12(m_u+m_d)=\Mn/3$ in this case. 
For the second moment (\ref{non-rel-limit}) yields $\int\di x\,e^q(x)=N_q/3$,
which is the correct non-relativistic result for the sum rule (\ref{2moment})
recalling that $m_q/\Mn=1/3$.

Thus, the non-relativistic result (\ref{non-rel-limit}) satisfies 
the QCD sum rules (\ref{1moment},\ref{2moment}).
However, the results $\sigmaPiN=\Mn$ and $\int\di x\,(e^u+e^d)(x)=3$ 
strongly overestimate and underestimate, respectively, the phenomenological 
numbers in Eqs.~(\ref{sigmaPiN}) and (\ref{1moment-number}).
In particular, one could be worried that such a large value for the pion
nucleon sigma term, $\sigmaPiN=\Mn$, would imply a huge number for the 
strangeness content $y$ of the nucleon, defined as
\be\label{y-1}
	y = \frac{2 \la N|\,\bar{\psi}_s\psi_s|N\ra}{\la N|\,
	(\bar{\psi}_u\psi_u+\bar{\psi}_d\psi_d)\,|N\ra}  \;. \ee
The precise role of $y$ in estimating the contribution of the strange quark 
degree of feedom to the nucleon mass was discussed in \cite{Ji:1994av}. 
In spite of the large strangeness content $y$ the total strange quark 
contribution to the nucleon mass is rather small \cite{Ji:1994av}. 

To leading order in chiral perturbation theory the relation between
$y$ and $\sigmaPiN$ is given by \cite{sigma-review}
\be\label{y-2}
	y = 1 - \frac{m}{m_s-m}\,\frac{\M{\Xi}+\M{\Sigma}-2\Mn}{\sigmaPiN}\;.
\ee
(With $m_s/m\simeq 25$ in Eq.~(\ref{y-2}) one obtains 
$y=1-26\,{\rm MeV}/\sigmaPiN$. Improved calculations in higher orders of 
chiral perturbation theory yield $y= 1-(35\pm5)\,{\rm MeV}/\sigmaPiN$ 
\cite{Gasser:1982ap}. However, for our purposes the relation (\ref{y-2}) 
is sufficient.)
Inserting the non-relativistic mass relations $\M{\Xi}=2m_s+m_q$, 
$\M{\Sigma}=m_s+2m_q$ and $\Mn = 3m_q$ into (\ref{y-2}) one observes that 
the mass of the strange quarks cancels out exactly, and $y=1-\Mn/\sigmaPiN$,
i.e.\ $y=0$ with the non-relativistic result $\sigmaPiN=\Mn$.
This result follows directly from (\ref{y-1}) since in the non-relativistic 
limit $\bar{\psi}_s\psi_s=\psi_s^\dag\psi_s$ is the number operator for 
strange quarks, which has a zero expectation value in the nucleon states.

Finally we observe that in the non-relativistic limit $e^q(x)$ is due to the
mass term in (\ref{e-decompostion-1}) only, as one intuitively would expect.
This can be seen by observing that 
$e^q_{\rm mass}(x) = (m_q/\Mn)\,f_1^q(x)/x = f_1^q(x)$,
since the $\delta(x-\frac13)$-function in (\ref{non-rel-limit}) 
allows to replace $x$ in the denominator by $\frac13$ which 
cancels the prefactor $m_q/\Mn=\frac13$. 
Thus, the non-relativistic ``description'' of $e^q(x)$ is theoretically 
consistent but phenomenologically not correct and, of course, not suited
to provide insights into the twist-3 structure of $e^q(x)$.

\paragraph{Bag model.}
The first model studies of $e^q(x)$ have been done
in the bag model \cite{Jaffe:1991ra,Signal:1997ct}.
In Fig.~\ref{Fig:bagCQSM} the results from Ref.~\cite{Jaffe:1991ra} are shown.
At the low scale of that model estimated to be around $0.4\,{\rm GeV}$
the quark distribution $e^q(x)$ is of comparable magnitude as $f_1^q(x)$. 
(The bag model is not expected to consistently describe antiquark 
distribution functions, since it yields $f_1^{\bar q}(x)<0$ in
contradiction with the positivity requirement.)

In the bag model the twist-2 Soffer inequality (\ref{Sof-twist-2}) is 
saturated,\footnote{
	This naturally happens in quark models with (sufficiently) 
	simple nucleon wave functions \cite{Goldstein:1995ek}.}
which is a necessary (but not sufficient) condition for the (generally 
incorrect) ``twist-3 inequality'' of Ref.~\cite{Soffer:1995ww} to be valid. 
Indeed, it is observed that the ``twist-3 Soffer-inequality'' 
holds in the bag model and is saturated, i.e.\ 
$e^q(x)= 2g_T^q(x) - h_L^q(x)$ in the bag model \cite{Signal:1997ct}.\footnote{
	This in turn is a necessary condition for the (also generally 
	incorrect) twist-4 inequality of Ref.~\cite{Soffer:1995ww} to be 
	valid, which again holds in the bag model and is saturated 
	\cite{Signal:1997ct}. 
	Thus, out of the three linearly independent quark distribution 
	functions, which exist in general at each twist level in a spin 
	$\frac12$ hadron, in the bag model only two respectively are linearly
 	independent.  In some sense this is analog to the situation in the 
	non-relativistic model, where $h_1^q(x)=g_1^q(x)$. 
	A possible explanation could be that both models contain only quarks
	and the longitudinally and transversely polarized nucleon states are 
	related to each other geometrically, namely by (respectively
	ordinary and Melosh) rotations.}

There is no $\delta(x)$-contribution to $e^q(x)$ in the bag model. 
The first moment arises from a valence-like structure and is of the order of 
magnitude of unity, underestimating the phenomenologically expected number in 
Eq.~(\ref{1moment-number}).  The sum rule for the second moment in 
Eq.~(\ref{2moment}) is violated in the bag model. 
This can be understood because this sum rule follows from equations 
of motion, which are modified by the bag boundary \cite{Jaffe:1991ra}.
It is worthwhile mentioning that $e^q(x)$ is entirely a bag surface effect.
This in some sense is consistent as the bag models confinement
and thus mimics gluons \cite{Jaffe:1991ra}.

\paragraph{Spectator model.}
The quark distribution functions $e^u(x)$ and $e^d(x)$ were estimated
in the spectator model at a scale pressumably lower than $0.5\,{\rm GeV}$ 
\cite{Jakob:1997wg}. In this model $e^u(x)$ was found sizeable and $e^d(x)$
rather small; i.e.\ $(e^u+e^d)(x)$ and $(e^u-e^d)(x)$ are of comparable 
magnitude, and the large-$N_c$ relation (\ref{large-Nc-2}) does not hold 
(which for finite $N_c$, of course, does not need to be the case).
Also in the spectator model there is no $\delta(x)$-function in $e^q(x)$.
The result for $(e^u+e^d)(x)$ is shown in Fig.~\ref{Fig:bagCQSM}a.

%
\FIGURE{
\begin{tabular}{cc}
\includegraphics[width=5cm,height=6cm]{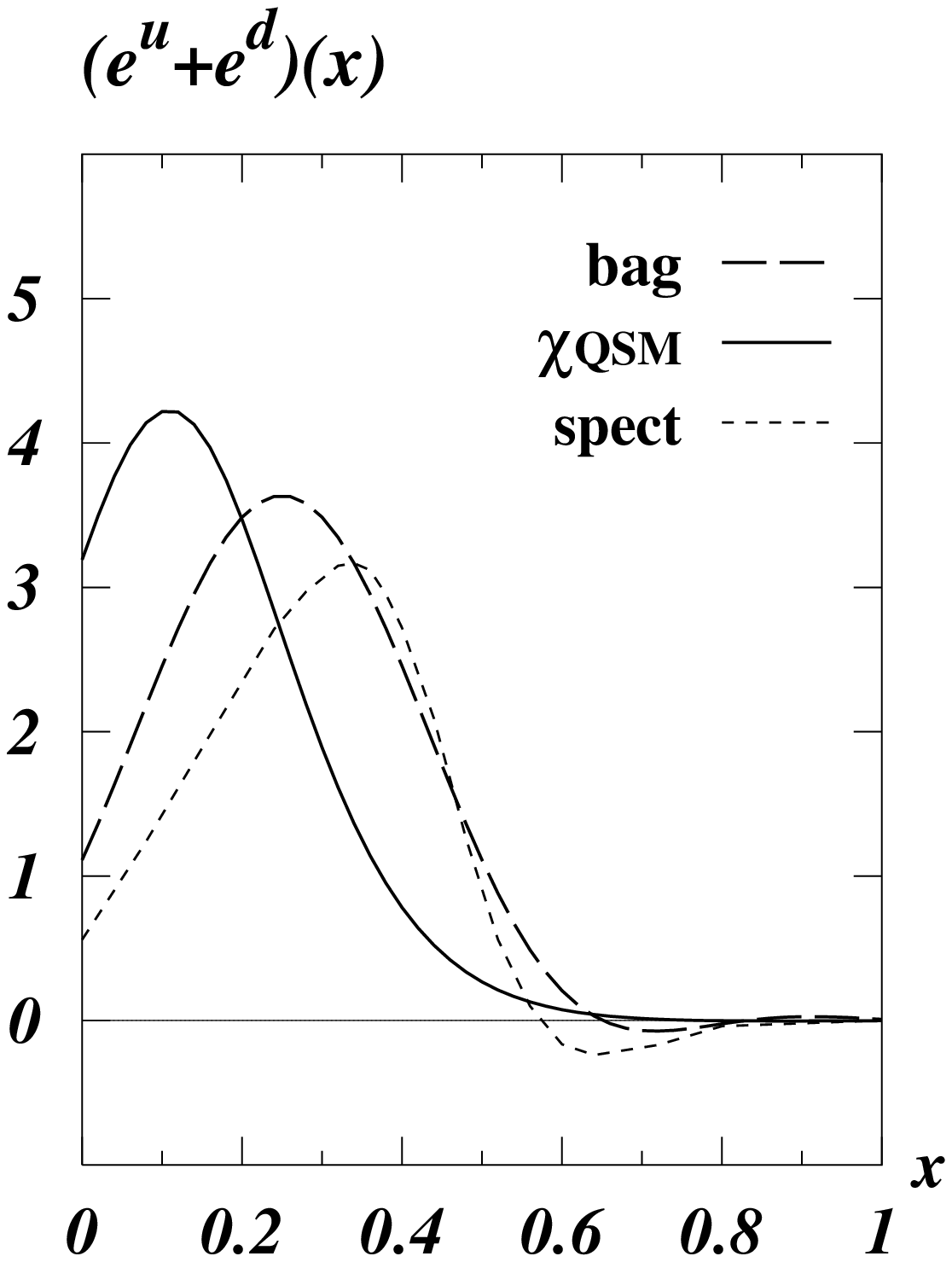}  & 
\includegraphics[width=5cm,height=6cm]{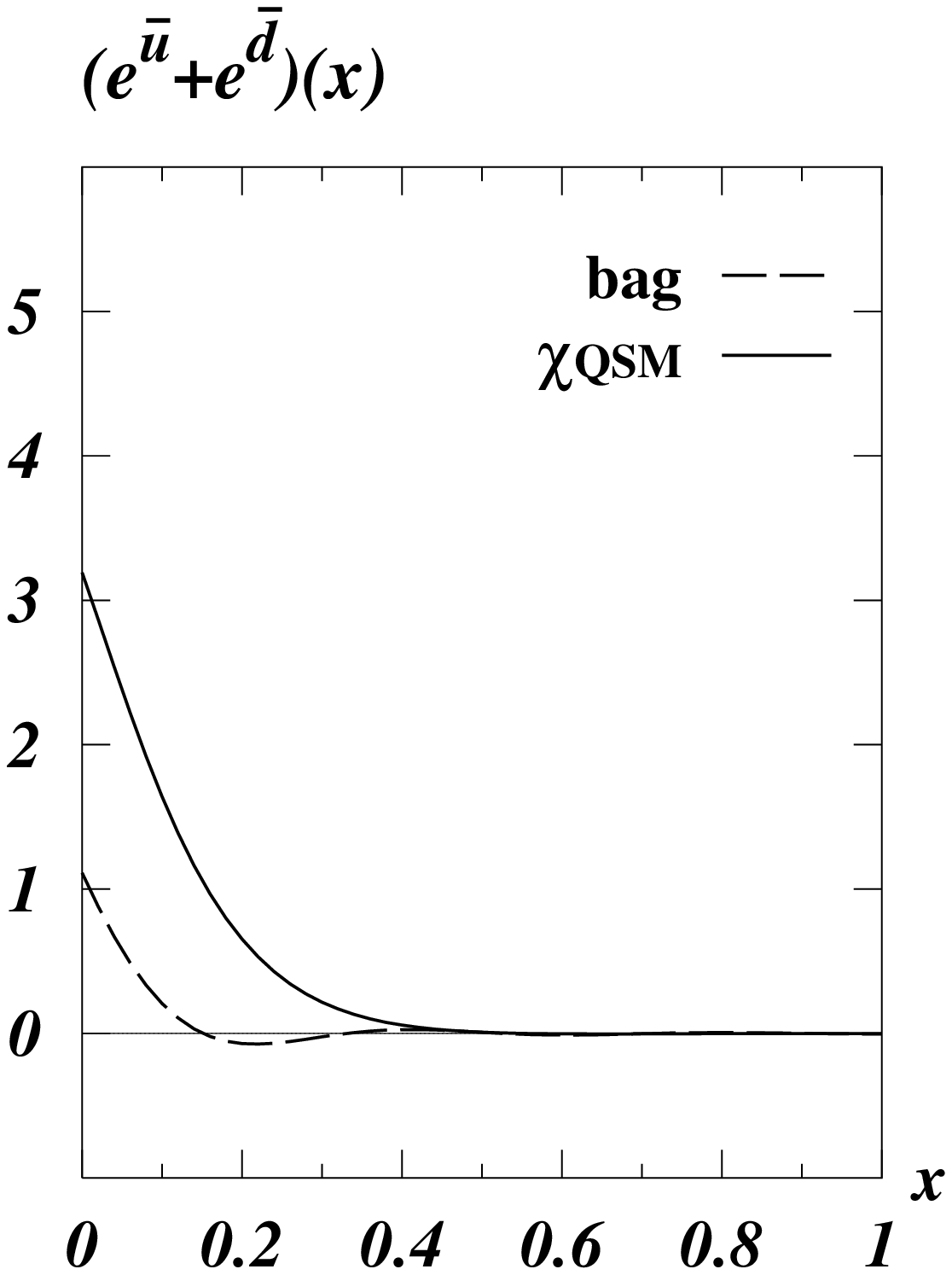}  \cr 
{\bf a} & 
{\bf b} 
\end{tabular}
	\label{Fig:bagCQSM}
	\caption{
	{\bf a} Results for the flavour singlet $(e^u+e^d)(x)$ vs.\ $x$
	from the bag \cite{Jaffe:1991ra}, spectator \cite{Jakob:1997wg} and 
	chiral quark-soliton model \cite{Schweitzer:2003uy}. 
	{\bf b} The same for anti-quarks.
	No attempt is made to indicate the $\delta(x)$-contribution in the 
	chiral quark-soliton model result. The model results refer to low 
	scales around $0.5\,{\rm GeV}$, see text.
	(The flavour-independent result for ``$e(x)$'' from 
	\cite{Jaffe:1991ra} is multiplied by a factor of $3$ for sake of 
	comparison to the spectator and chiral quark-soliton model results.)}}
%

\paragraph{Gross-Neveu model in $(1+1)$-dimensions.}
In Ref.~\cite{Burkardt:1995ts} $e^a(x)$ was discussed in several toy models.
The purpose of these studies was not to provide realistic estimates of 
$e^a(x)$, but to shed some light on possible mechanisms leading to the 
appearance of a $\delta(x)$-term. 
In particular, in a {\sl non-perturbative} calculation in the
$(1+1)$-dimensional version of the Gross-Neveu model the twist-3 distribution 
of ``s''-type (model-) quarks in the ``u''-type (model-) quark 
$e^{s/u}(x)=\frac{1}{2M}\int\frac{\di\lambda}{2\pi}e^{i\lambda x}
\la u|\bar{\psi}_s(0)\psi_s(\lambda n)|u\ra$ was found to be proportional
to $\delta(x)$. In the Gross-Neveu model the $\delta(x)$-term arises 
from long range correlations along the light cone, i.e.\  from the 
fact that the correlator $\la u|\bar\psi_s(0)\psi_s(z)|u\ra$ ($z$ light-like) 
contains a contribution independent of $z$ \cite{Burkardt:1995ts}. In order 
to illustrate that this need not be a peculiarity of the $(1+1)$-dimensional 
model it was shown (by means of {\sl perturbative} methods) that $e^a(x)$ 
contains a $\delta(x)$-contribution also in other $(1+1)$- and 
$(3+1)$-dimensional field theories \cite{Burkardt:1995ts} 
(cf.\ also \cite{Burkardt:2001iy}).

\paragraph{\boldmath Chiral quark-soliton model ($\chi$QSM).}
The flavour-singlet combination $(e^u+e^d)(x)$, which is the leading 
contribution in the large-$N_c$ limit (\ref{large-Nc-behaviour-in-general}), 
has been studied in the $\chi$QSM at a low normalization point of about
$0.6\,{\rm GeV}$ \cite{Schweitzer:2003uy}. 
Interestingly, $(e^u+e^d)(x)$ was found to contain a $\delta(x)$-contribution
\cite{Schweitzer:2003uy}
\be\label{e-CQSM-1}
	(e^u+e^d)(x) = C\,\delta(x) + (e^u+e^d)(x)_{\rm reg} \;, \ee
where $(e^u+e^d)(x)_{\rm reg}$ is a regular part, which has a valence-like 
structure and qualitatively looks similar to the bag and spectator model 
results, see Fig.~\ref{Fig:bagCQSM}.
The coefficient $C$ is quadratically UV-divergent in the $\chi$QSM and can 
consistently be regularized. It is remarkable that in the model the baryonic 
quantity $C$ is proportional to the quark vacuum condensate
\be\label{e-CQSM-2}
	C = A_N\,\la{\rm vac}|(\bar\psi_u\psi_u+\bar\psi_d\psi_d)|{\rm vac}\ra
	\;,\;\;\;  A_N = \frac12\int\!\di^3{\bf x}\;
	\trF\biggl(\frac{U+U^\dag}{2}-1\biggr) \;. \ee
The proportionality factor $A_N$ encodes the information on the nucleon:
$U=\exp(i\tau^a\pi^a)$ is the SU(2) chiral pion (soliton) field, 
$\trF$ denotes the trace over flavour. Numerically $C = (9\pm 3)$ 
for the value $(-280\pm 30)^3{\rm MeV}^3$ of the quark-condensate 
from Ref.~\cite{Gasser:1982ap}.
In the $\chi$QSM the first moment of $(e^u+e^d)(x)$ is not solely due to 
the $\delta(x)$-function but also receives a (small) contribution from
the regular part $(e^u+e^d)(x)_{\rm reg}$ in (\ref{e-CQSM-1}).
The final results are $\int\di x\,(e^u+e^d)(x) = (10\pm3)$ and
$\sigmaPiN= 80\;{\rm MeV}$ 
in reasonable agreement with Eqs.~(\ref{sigmaPiN},\ref{1moment-number}).
The sum rules for the first and the second\footnote{
	For the second moment, however, one cannot expect the current quark
	mass $m_q$ to appear in the model-version of the QCD sum rule 
	(\ref{2moment}). Instead an ``effective'' mass appears because
	the $\chi$QSM describes {\sl bound} quarks at a {\sl low scale} 
	of $0.6\,{\rm GeV}$.}
moment are satisfied in the $\chi$QSM.
There are no means in this model (such as gauge-invariance in QCD) to further
decompose the regular part in (\ref{e-CQSM-1}), which is to be understood as
the entangled pure-twist-3-term and mass-term.

In Ref.~\cite{Wakamatsu:2003uu} the existence of a $\delta(x)$ contribution 
in $(e^u+e^d)(x)$ in the $\chi$QSM was concluded independently and in an 
alternative way to the derivation given in \cite{Schweitzer:2003uy}.
In particular, in Ref.~\cite{Wakamatsu:2003uu} it was shown that the 
$\delta(x)$-term is due to long-distance quark-quark correlations.
Thus, the underlying non-perturbative mechanism which gives rise to the
$\delta(x)$ contribution in the $\chi$QSM is analog to that 
observed in the Gross-Neveu model \cite{Burkardt:1995ts}.

Apparently, a $\delta(x)$-contribution has no partonic interpretation. 
However, the model relation (\ref{e-CQSM-2}) ``suggests'' an 
``intuitive understanding''.
Let us simplifyingly interpret $e^a(x)$ as scattering (in a particular way) 
off a parton in the nucleon, which carries the nucleon momentum fraction $x$
(in the infinite momentum frame) \cite{Schweitzer:2003uy}. 
What means scattering off a parton, which carries the momentum fraction $x=0$? 
Eq.~(\ref{e-CQSM-2}) suggests that the parton with $x = 0$ is not ``moving'' 
with the fast proton but indeed at rest. And it is not taken out of the proton
but out of the vacuum, which to some extent (quantified by the constant 
$A_N$ in (\ref{e-CQSM-1})) is present also inside the proton 
\cite{Schweitzer:2003uy}. 
It would be interesting to see whether the naive ``interpretation'' could be 
``confirmed'' by observing relations analogue to (\ref{e-CQSM-2}) in other 
chiral models.

Finally we remark that the coefficient $C$ of the $\delta(x)$-contribution
in Eq.~(\ref{e-CQSM-2}) implies a relation between the pion nucleon sigma 
term and the quark vacuum condensate 
\be\label{sigmaPiN-vac-cond}
	\sigmaPiN = m\,
	\la{\rm vac}|(\bar\psi_u\psi_u+\bar\psi_d\psi_d)|{\rm vac}\ra \, A_N 
	\ee
with $A_N$ (which is negative) as defined in Eq.~(\ref{e-CQSM-2}). 
The only model in which such a relation was known so far is the 
Skyrme model \cite{Adkins:1983hy}.\footnote{
	We thank the referee for pointing out this to us.}

\paragraph{Summary of model results.} 
In the non-relativistic limit $e^a(x)$ and $f_1^a(x)$ become equal. 
The more realistic bag, spectator and chiral quark-soliton model
\cite{Jaffe:1991ra,Schweitzer:2003uy,Signal:1997ct,Jakob:1997wg} suggest 
that $e^a(x)$ has a sizeable valence-like structure at a low scale
and is roughly half the magnitude of $f_1^a(x)$.
The equations of motions are modified in these models (compared to QCD)
and there is no gauge principle. Therefore a decomposition analogue to 
(\ref{e-decompostion-1}) is not possible.
Still, in the $\chi$QSM there is a $\delta(x)$-contribution.

The model results certainly do not discourage measurements of observables 
containing information on $e^q(x)$. However, as will be discussed in the
next section, this is a difficult task and only recently progress has
been reported.

\section{\boldmath $e^q(x)$ in experiments}
\label{Sec:e-in-experiments}

The distribution function $e^a(x)$ is a ``spin-average'' distribution, 
i.e.\ accessible in experiments with unpolarized nucleons. However,
due to its chiral odd nature it can enter an observable only in connection 
with another chirally odd distribution or fragmentation function, and 
due to twist-3 its contribution is suppressed by a factor of $\Mn/Q$, 
where $Q$ denotes the hard scale of the process.
E.g. the combination $\sum_q e_q^2\;e^q(x)\,e^{\bar q}(x)$ contributes to 
the Drell-Yan process with unpolarized proton beams, but only at twist-4 
and together with other twist-4 quark-gluon-correlation functions
\cite{Jaffe:1991kp,Jaffe:1991ra}. 
For some time this was the only known process involving $e^a(x)$,
which of course is inpractical to access $e^a(x)$ experimentally.

Then the chirally and T-odd ``Collins fragmentation function'' 
$H_1^{\perp a}(z)$ has been introduced \cite{collins,muldt}, which describes 
the left-right asymmetry in the fragmentation of a transversely polarized 
quark of flavour $a$ into a hadron.  $H_1^{\perp a}(z)$ is ``twist-2'' in 
the sense that its contribution to observables is not power suppressed.
Assuming factorization for transverse momentum dependent processes, 
it was shown that the combination $\sum_q e_q^2\;e^q(x)\,H_1^{\perp q}(z)$
gives the dominant (tree-level) contribution to a single spin asymmetry, 
$A_{LU}^{\sin\phi}$, in hadron production from semi-inclusive DIS of 
longitudinally (subscript $_L$) polarized electrons off unpolarized 
(subscript $_U$) protons \cite{muldt,Levelt:1994np,Mulders:1996dh}.
$A_{LU}^{\sin\phi}$ is proportional to the sine of the azimuthal 
angle $\phi$ of the produced pion around the z-axis defined by the 
exchanged virtual photon.
This azimuthal asymmetry has been measured in the HERMES experiment and 
found consistent with zero within error bars \cite{hermes,hermes-pi0}.
More recently, however, the CLAS collaboration reported the measurement of 
a non-zero $A_{LU}^{\sin\phi}$ in a different kinematics 
\cite{Avakian-talk,Avakian:2003pk}.

%
\FIGURE{
\includegraphics[width=5.9cm,height=6.2cm]{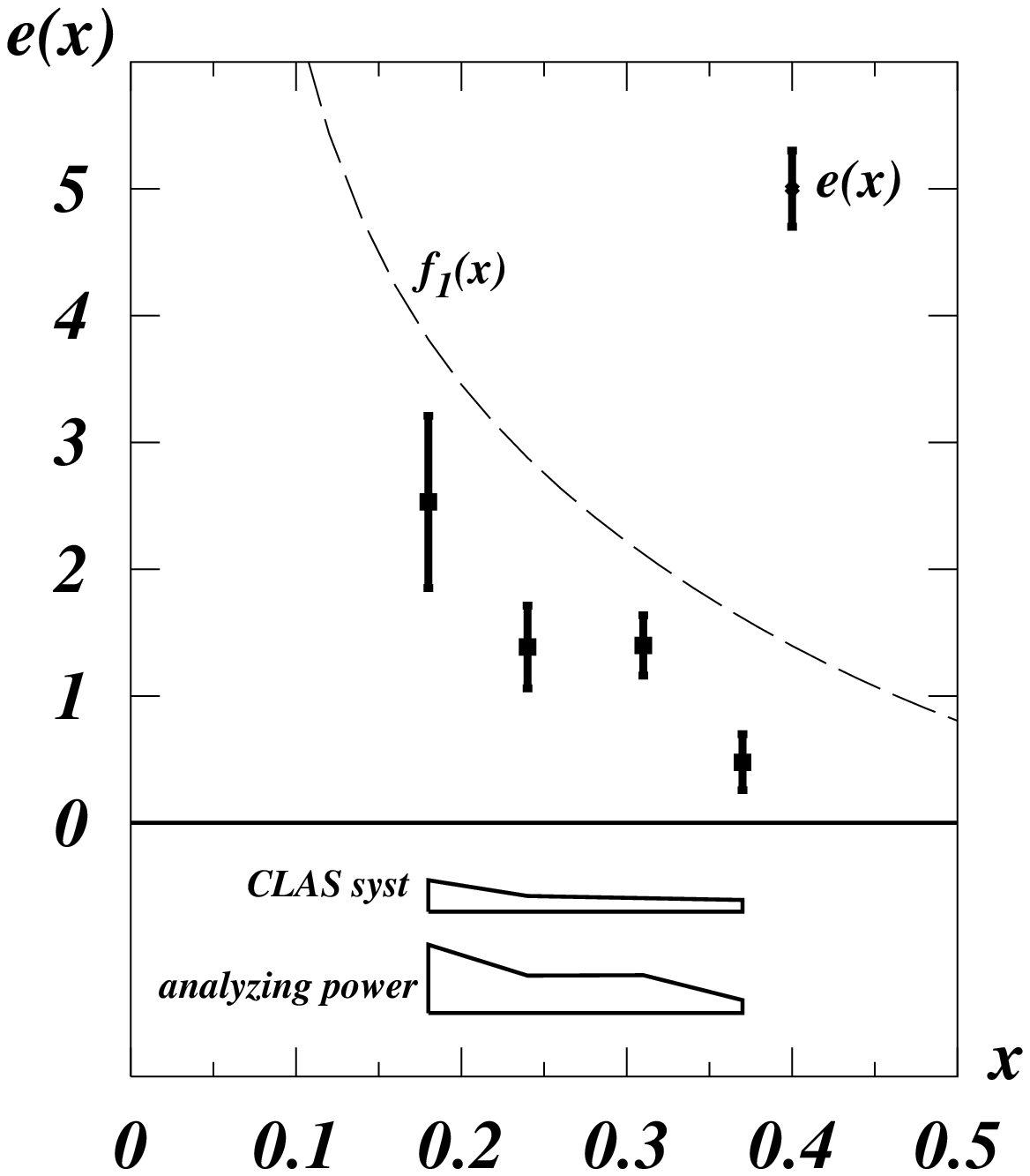}  
	\label{Fig:CLAS}
	\caption{The $e(x)\equiv(e^u+\frac14 e^{\bar d})(x)$ at 
	$\la Q^2\ra=1.5 \,{\rm GeV}^2$ vs.\  $x$ as extracted in 
	\cite{Efremov:2002ut} from the CLAS data \cite{Avakian:2003pk}.
	For comparison is shown the corresponding flavour-combination of 
	$f_1^a(x)$ (from \cite{Gluck:1994uf}).}}
%

In the HERMES experiments also another azimuthal asymmetry, 
$A_{UL}^{\sin\phi}$, in pion production from semi-inclusive DIS of unpolarized
positrons off a longitudinally polarized proton target has been studied 
and found to be sizeable for $\pi^+$ and $\pi^0$ \cite{hermes,hermes-pi0}.
This asymmetry contains information on $H_1^{\perp a}(z)$ and the chirally 
odd twist-2 $h_1^a(x)$ and twist-3 $h_L^a(x)$ distribution functions 
\cite{Mulders:1996dh}.
Using the model predictions from Refs.~\cite{h1-model,Dressler:2000hc} for 
$h_1^a(x)$ and $h_L^a(x)$, in Ref.~\cite{Efremov:2001cz} the relation
among the favoured Collins and unpolarized fragmentation functions
$H_1^{\perp}(z) = (0.33\pm 0.06)\, z\,D_1(z)$ for $0.2\le z\le 0.7$ at 
$\la Q^2\ra=2.4\,{\rm GeV}^2$ has been extracted from the HERMES data 
\cite{hermes,hermes-pi0}.

This result for $H_1^\perp(z)$ has been used in Ref.~\cite{Efremov:2002ut} 
to extract first experimental information on the flavour combination 
$(e^u+\frac14 e^{\bar d})(x)$ from the CLAS data on $A_{LU}^{\sin\phi}$ 
in $\pi^+$ production \cite{Avakian:2003pk}. 
For that it was assumed that the CLAS data is due to the Collins effect
and the tree-level description of the process of 
Refs.~\cite{Levelt:1994np,Mulders:1996dh} can be applied at the modest 
scale $\la Q^2\ra = 1.5 \,{\rm GeV}^2$ of the CLAS experiment. 
The extracted $(e^u+\frac14 e^{\bar d})(x)$ is definitely not small, about 
half the magnitude of the corresponding flavour combination of $f_1^a(x)$ 
in the region $0.15\le x\le 0.4$ covered in the CLAS experiment,
see Fig.~\ref{Fig:CLAS}. 
It will be interesting to see whether the first information on $e^a(x)$ 
reported in Ref.~\cite{Efremov:2002ut} will be confirmed in future 
experiments.

The process described above is at present the cleanest way to access $e^a(x)$.
However, there are other processes, e.g.\  electro-production of transversely 
polarized $\Lambda$ from SIDIS of a longitudinally polarized electron beam off
an unpolarized proton target, where $e^a(x)$ contributes together with further
unknown fragmentation and distribution functions \cite{Mulders:1996dh}.

\section{Conclusions}
\label{Sec:conlusions}

A brief discussion of the twist-3 chirally odd distribution function $e^a(x)$ 
has been given, focusing on theoretical properties, model estimates and 
perspectives to measure $e^a(x)$.

In particular, it has been emphasized that QCD equations of motion imply the
existence of a $\delta$-type singularity in  $e^a(x)$ at $x=0$. The first 
Mellin moment of $e^a(x)$ is solely due to this $\delta(x)$-contribution. 
This means that unfortunately DIS-experiments will not provide any information
on the phenomenologically interesting pion nucleon sigma-term, which formally
is related to the first moment of $e^a(x)$. 
Historically these conclusions have been drawn first on the basis 
of Regge arguments before the advent of QCD \cite{Jaffe:1972yv}. 

The existence of the $\delta(x)$-function, however, can indirectly be 
confirmed in the experiment by observing that the first moment of $e^a(x)$ 
vanishes if the point $x=0$ is not included in the $x$-integration. 
If current quark mass effects are neglected (which practically are suppressed 
by $m_q/Q$), $e^a(x)$ satisfies the following ``practical DIS-sum rules''
\be\label{final}
	\int\limits_{0\!+}^1\!\!\di x\;(e^q+e^{\bar q})(x) \approx 0\;, \;\;\;
	\int\limits_0^1\!\!  \di x\;x\,(e^q-e^{\bar q})(x) \approx 0\;. \ee
The integration limit $0\!+$ in the sum rule for the first moment means 
that the point $x=0$ is not included.  This corresponds to the experimental 
situation, since data can be obtained only for $x\ge x_{\rm min}>0$, with 
$x_{\rm min}$ depending on the facility.
In principle the vanishing of the second moment of $e^a(x)$ 
could also be tested experimentally by taking the difference between
``structure functions'' $\sum_a e_a^2\,e^a(x)$ extracted from proton and 
neutron data, and assuming a flavour symmetric sea, i.e.\ the relation 
$\int_0^1\di x\,x(e^{\bar u}-e^{\bar d})(x)=0$.
(However, in the case of $f_1^{\bar q}(x)$ an analogue relation, 
the ``Gottfried sum rule'' turned out to be wrong:
$\int_0^1\di x\,(f_1^{\bar u}-f_1^{\bar d})(x) \neq 0$ \cite{Kumano:1997cy}.)

The ``practical'' sum rules (\ref{final}) recall the situation of the 
twist-3 distribution function $\overline{g}_2(x)=g_2(x)-g_2^{WW}(x)$.
The first moment of $\overline{g}_2(x)$ vanishes due to the
Burkhardt-Cottingham sum rule \cite{Burkhardt:ti} and the second moment 
due to the Efremov-Teryaev-Leader-sum rule \cite{Efremov:1996hd}.
In the case of the chirally odd distribution $e^a(x)$ the sum rules
(\ref{final}) will be even more difficult to test in the experiment.

In the limit of a large number of colours $N_c$ one finds 
$e^u(x) = e^d(x)$ modulo $1/N_c$ corrections, and in the non-relativistic 
one obtains $e^q(x) = f_1^q(x)$ modulo relativistic corrections.
Both relations could serve as useful guidelines.
In relativistic models, such as the bag \cite{Jaffe:1991ra,Signal:1997ct},
spectator \cite{Jakob:1997wg} and chiral quark-soliton model ($\chi$QSM) 
\cite{Schweitzer:2003uy}, $e^a(x)$ has a sizeable valence-like structure at 
low scales about $0.5\,{\rm GeV}$ and is roughly half the magnitude of 
$f_1^a(x)$.
These models do not respect the practical DIS sum rules (\ref{final})
since the latter follow from the QCD equations of motions -- but in the bag 
and chiral quark-soliton model different (model-) equations of motions hold.
Still, in the $\chi$QSM there is a $\delta(x)$ in $e^q(x)$
\cite{Schweitzer:2003uy,Wakamatsu:2003uu}. 
The $\delta(x)$-function arises in the $\chi$QSM from long-distance 
quark-quark correlations \cite{Wakamatsu:2003uu} -- i.e.\  from basically the
same non-perturbative mechanism found previously in the Gross-Neveu model 
in $(1+1)$-dimensions \cite{Burkardt:1995ts}. 

Experimentally $e^a(x)$ could be accessed by means of the Collins effect 
\cite{collins} in semi-inclusive DIS of polarized electrons off an 
unpolarized target \cite{Levelt:1994np,Mulders:1996dh}. 
Recently the CLAS collaboration \cite{Avakian-talk,Avakian:2003pk} 
studied the process $\vec{e}p\to\pi^+X$ and observed a particular 
angular distribution of the produced pions in the single (beam) spin 
asymmetry -- as one would expect on grounds of the Collins effect
\cite{collins,Levelt:1994np,Mulders:1996dh}.
If this interpretation, that the CLAS data \cite{Avakian-talk,Avakian:2003pk}
are due to the Collins effect, applies then $e^a(x)$ is definitely 
not small, roughly half the magnitude of $f_1^a(x)$ at a scale of 
about $1.5\,{\rm GeV}^2$ \cite{Efremov:2002ut}.

The CLAS experiment could provide further insights, as could possibly do 
other fixed target experiments such as HERMES and COMPASS -- when focusing 
on the large $z$ region where the analyzing power $H_1^\perp(z)/D_1(z)$
is larger \cite{Efremov:2001cz}.
Also EMC at CERN (by a reanalysis of old data) and HERA at DESY 
could provide further information -- where the advantage of a polarized 
lepton beam and an unpolarized proton could be used to explore especially 
the small $x$-region at moderate $Q^2$ needed to resolve the twist-3 effect.

\acknowledgments

We are grateful to G.~Altarelli and A.~V.~Belitsky for discussions, and 
to the Department of Theoretical Physics of Turin University and CERN
for warm hospitality where parts of this work have been completed. We also 
would like to thank the referee for several interesting and valuable remarks.
A.~E.\ is partially supported by RFBR grant 03-02-16816
and INTAS grant 00/587.
This work has partly been performed under the contract  
HPRN-CT-2000-00130 of the European Commission.

\appendix
\section{Consequences from using equations of motions}
\label{App:using-eq-of-motion}

In this Appendix the consequences are explored of using the identity 
in Eq.~(\ref{identity}) based on the equations of motions.
The Mellin moments are defined as 
${\cal M}_n[q]=\int_{-1}^1\di x\,x^{n-1} q(x)$.

\paragraph{\boldmath The $\delta$-function.} 
The singular contribution originates from the local term 
$\bar\psi_q(0)\psi_q(0)$ in the identity (\ref{identity}) and is given by
\be\label{App-sing-1}
	e^q_{\rm sing}(x) 
	 = \frac{1}{2\Mn} \int\!\frac{\di\lambda}{2\pi}\,e^{i\lambda x}
	   \;\la N|\bar\psi_q(0)\psi_q(0)|N\ra
	 = \delta(x)\;\frac{1}{2\Mn}\;\la N|\bar\psi_q(0)\psi_q(0)|N\ra\;,\ee
with the Mellin moments ${\cal M}_n[e^q_{\rm sing}]=$
$\delta_{n1}\,\frac{1}{2\Mn}\;\la N|\bar\psi_q(0)\psi_q(0)|N\ra$.

\paragraph{The pure twist-3 term.} 
The second term on the right hand side (RHS) of the identity 
(\ref{identity}) gives rise to
\ba\label{App-twist3-1}
&&	e^q_{\rm tw3}(x) 
	= \frac{1}{4\Mn} 
	\int\!\frac{\di\lambda}{2\pi}\,e^{i\lambda x}\,\lambda^2\;
	{\cal F}^q_{\rm tw3}(\lambda) \;, \\
&&	{\cal F}^q_{\rm tw3}(\lambda) =
	\int\limits_0^1\di u\int\limits_0^u\di v\;\la N|\,
	\bar\psi_q(0)\,\sigma^{\alpha\beta}\,n_\beta\,[0,v\lambda n]\,
	gG_{\alpha\nu}(v\lambda n)\,n^\nu\,[v\lambda n,u\lambda n]\,
	\psi_q(u\lambda n) \,|N\ra \;.\;\;\;\nonumber
\ea
For the Mellin moments one obtains (using the
support property $e^a(x) \equiv 0$ for $|x|\ge 1$)
\ba\label{App-twist3-2}
&&	\mbox{\hspace{-0.3cm}}
	{\cal M}_n[e^q_{\rm tw3}] 
	= \frac{1}{4\Mn}
	\int\limits_{-\infty}^\infty\!\!\di x \; x^{n-1} 
	\int\limits_{-\infty}^\infty\!\frac{\di\lambda}{2\pi}\,e^{i\lambda x}
	\,\lambda^2\; {\cal F}^q_{\rm tw3}(\lambda)  \\
&&	\mbox{\hspace{-0.3cm}} = 
	\frac{1}{4\Mn}\int\limits_{-\infty}^\infty\!\di\lambda\;
	\,\lambda^2\;{\cal F}^q_{\rm tw3}(\lambda) 
	\biggl(\frac{\partial\;\;}{i\partial\lambda}\biggr)^{\!\!n-1}
	\int\limits_{-\infty}^\infty\!\!\frac{\di x}{2\pi}\; e^{i\lambda x} 
	= \frac{1}{4\Mn}
	\biggl(-\frac{\partial\;\;}{i\partial\lambda}\biggr)^{\!\!n-1}
	\,\lambda^2\;{\cal F}^q_{\rm tw3}(\lambda) 
	\biggl|_{\lambda = 0} \;.\nonumber \ea
From Eq.~(\ref{App-twist3-2}) we immediately see that the 
first two moments of $e^q_{\rm tw3}(x)$ vanish
\be\label{App-twist3-3}
	{\cal M}_1[e^q_{\rm tw3}] = 
	{\cal M}_2[e^q_{\rm tw3}] = 0 \;. \ee
Higher Mellin moments, $n>2$, are generally non-zero. For explicit 
expressions see \cite{Balitsky:1996uh,Belitsky:1997zw,Koike:1997bs}.

\paragraph{The mass term.}
The mass term follows from the third operator on the RHS of the identity
(\ref{identity})
\be\label{App-mass-term-1}
	e^q_{\rm mass}(x) = -\,\frac{m_q }{\Mn} 
	\int\!\frac{\di\lambda}{4\pi}\,e^{i\lambda x}\,i\lambda
	\int\limits_0^1\di u\; \la N|\,\bar\psi_q(0)\not{\!n}\,[0,u\lambda n]
	\,\psi_q(u\lambda n)|N\ra \;.\ee
Taking Mellin moments of $e^q(x)^{\rm mass}$ (and using the
support property) yields
\ba\label{App-mass-term-1a}
&&	{\cal M}_n[e^q_{\rm mass}] 
	= 
	-\,\frac{m_q }{\Mn} \int\limits_{-\infty}^\infty\!\!\di x \; x^{n-1}
	\int\limits_{-\infty}^\infty\!\frac{\di\lambda}{4\pi}\,
	i\lambda\,e^{i\lambda x}
	\int\limits_0^1\di u\; \la N|\,\bar\psi_q(0)\not{\!n}\,[0,u\lambda n]
	\,\psi_q(u\lambda n)|N\ra \nonumber\\
	&&= -\,
	\frac{m_q }{2\Mn}\int\limits_{-\infty}^\infty\!\di\lambda\,i\lambda\,
	\int\limits_0^1\di u\; \la N|\,\bar\psi_q(0)\not{\!n}\,[0,u\lambda n]
	\,\psi_q(u\lambda n)|N\ra 
	\biggl(\frac{\partial\;\;}{i\partial\lambda}\biggr)^{\!n-1}
	\int\limits_{-\infty}^\infty\!\frac{\di x}{2\pi}\;e^{i\lambda x} 
	\nonumber\\
	&&= -\,
	\frac{m_q }{2\Mn}\int\limits_0^1\di u\;
	\biggl(-\frac{\partial\;\;}{i\partial\lambda}\biggr)^{\!\!n-1}
	i\lambda\, \la N|\,\bar\psi_q(0)\not{\!n}\,[0,u\lambda n]
	\,\psi_q(u\lambda n)|N\ra\Biggr|_{\lambda = 0} \nonumber\\
	&&=
	\cases{ 0 \phantom{\biggr|}& for $n=1$ 
	\cr
	\frac{m_q }{2\Mn}\int\limits_0^1\di u\;(n-1)
	(i\frac{\partial}{\partial\lambda})^{n-2}
	\la N|\,\bar\psi_q(0)\not{\!n}\,[0,u\lambda n]
	\,\psi_q(u\lambda n)|N\ra\biggr|_{\lambda = 0}
	& for $n\ge 2$.} \ea
Using the relation
$$
	\displaystyle
 	\biggl(i\frac{\partial}{\partial\lambda}\biggr)^{\!\!m} 
	[0,u\lambda n]\,\psi(u\lambda n)\biggr|_{\lambda = 0}
	= (i\,u\, n_\alpha D^\alpha)^m\psi(0) $$
in the expression for $n\ge 2$ yields
\ba	&&	
	\frac{m_q }{2\Mn}\int\limits_0^1\di u\;(n-1)\,
	\biggl(i\frac{\partial}{\partial\lambda}\biggr)^{\!\!n-2}
	\la N|\,\bar\psi(0)\not{\!n}\,[0,u\lambda n]
	\,\psi(u\lambda n)|N\ra\biggr|_{\lambda = 0} \nonumber\\
	&&=
	\frac{m_q }{2\Mn}\la N|\,\bar\psi(0)\not{\!n}\,
	(i\, n_\alpha D^\alpha)^{n-2}\,\psi(0)|N\ra
	\int\limits_0^1\di u\;(n-1)\,u^{n-2} \nonumber \ea
such that we obtain
\be\label{App-mass-term-2}
	{\cal M}_n[e^q_{\rm mass}] =
	\cases{ 0 \phantom{\Biggr|}& for $n=1$ \cr \displaystyle
	\frac{m_q }{2\Mn}\la N|\,\bar\psi(0)\not{\!n}\,
	(i\,n_\alpha D^\alpha)^{n-2}\,\psi(0)|N\ra
	& for $n\ge 2$.} \ee
Recalling the definition of  the twist-2 ``unpolarized''  distribution 
function $f_1^q(x)$ and its moments
\ba\label{App-mass-term-3}
&&	f_1^q(x) = \int\!\frac{\di\lambda}{4\pi}\,e^{i\lambda x}\,
	\la N|\,\bar{\psi}_q(0)\not{\!n}\,[0,\lambda n]\,\psi_q(\lambda n)
	\,|N\ra \;,\;\;\; f_1^{\bar q}(x)= - f_1^q(-x) \nonumber\\
&&	{\cal M}_n[f_1^q] = \frac12
	\la N|\,\bar{\psi}_q(0)\not{\!n}\,(i\,n_\alpha D^\alpha)^{n-1}
	\,\psi_q(0)\,|N\ra
\ea
we obtain the relation between the mass term and $f_1^q(x)$ 
quoted in Eq.~(\ref{e-decompostion-mass-mom}).

In the formal manipulations in 
Eqs.~(\ref{App-twist3-2},\ref{App-mass-term-1a}) it was assumed that 
the order of integrations over $x$ and $\lambda$ can be interchanged.
In general this may not be allowed (see e.g.\ the discussion in Sec.~5.4 
of Ref.~\cite{Anselmino:1994gn}). However, in above cases one does not
need to worry, because all moments of $e^a(x)$ are well defined 
(since $\sigmaPiN$, $N_q\frac{m_q}{\Mn}$, etc.\ are finite).


\end{document}